\documentstyle[12pt,a4,epsf]{article}
\begin{document}
\begin{titlepage}
\title{
\vspace*{-10ex}
\hfill
\parbox{40mm}
{\normalsize KUCP-0110\\OHU-9710\\UT-790\\quant-ph/9710064}\\
\vspace{1cm}
{\Large \bf Valleys in Quantum Mechanics}}
\author{Hideaki {\sc Aoyama}$^{\dagger}$,
Hisashi {\sc Kikuchi}$^{\dagger\dagger}$, 
Ikuo {\sc Okouchi}$^{\ddagger}$,
\\
Masatoshi {\sc Sato}$^{\star}$ 
and Shinya {\sc Wada}$^{\ddagger}$
\vspace{0.5cm}
\\
$^{\dagger}${\normalsize\sl Department of Fundamental Sciences,}\\ 
{\normalsize\sl Faculty of Integrated Human Studies,}
\\
{\normalsize\sl Kyoto University, Kyoto 606-01, Japan}
\\
{\normalsize\tt aoyama@phys.h.kyoto-u.ac.jp}
\vspace{0.2cm}
\\
$^{\dagger\dagger}${\normalsize\sl Ohu University, Koriyama 963, 
Japan}
\\
{\normalsize\tt kikuchi@yukawa.kyoto-u.ac.jp}
\vspace{0.2cm}
\\
$^{\ddagger}${\normalsize\sl Graduate School of Human and Environmental 
Studies,}
\\
{\normalsize\sl Kyoto University, Kyoto 606-01, Japan}
\\
{\normalsize\tt dai@phys.h.kyoto-u.ac.jp, shinya@phys.h.kyoto-u.ac.jp}
\vspace{0.2cm}
\\
$^{\star}${\normalsize\sl Department of Physics, 
University of Tokyo, Tokyo 113, Japan}
\\
{\normalsize\tt msato@hep-th.phys.s.u-tokyo.ac.jp}}
\date{\normalsize October, 1997}

\maketitle
\thispagestyle{empty}
\begin{abstract} 
\normalsize
Conventionally, perturbative and non-perturbative calcula-\break
tions 
are performed independently.  
In this paper, valleys in the configuration space in quantum mechanics
are investigated as a way to treat them in a unified manner.
All the known results of the interplay of them are reproduced
naturally. 
The prescription for separating the non-perturbative contribution
from the perturbative is given in terms of the analytic
continuation of the valley parameter.  
Our method is illustrated on a new series
of examples with the asymmetric double-well potential.
We obtain the non-perturbative part explicitly,
which leads to the prediction of the large order behavior of the 
perturbative series.
We calculate the first 200 perturbative coefficients 
for a wide range of parameters and confirm the agreement with 
the prediction of the valley method.

\end{abstract}
\end{titlepage}

\newpage

In this paper, we report on  ``valleys'' in the configuration space of 
the path-integral. The valley is a series of configurations which 
connects a classical minimum (vacuum) with an instanton-like configuration.
As will be shown in the following, it reveals how non-perturbative
effects interfere with perturbative effects.

The perturbative expansion is the most common method to analyze models
in quantum mechanics or quantum field theories.
Its predictive power is, however, in doubt in a strict sense,
for the perturbative series is known to be divergent in most of the quantum
systems \cite{Dys}.
Models in quantum mechanics have been intensively studied in order to
understand the problem and to obtain physical predictions \cite{review}.
The symmetric double-well potential provides a very useful 
testing ground in this respect \cite{BPZ}.
Although the energy spectrum must be real by definition, the naive
application of the Borel-sum method to the perturbation series results
in a complex spectrum.
The interplay of non-perturbative effect, the tunneling between the
wells, was inferred to resolve this complexity \cite{Bog,Zin}.
%The conjecture inferred in Ref.\cite{Zin} is that the imaginary parts
%by the Borel summation is canceled by those induced in 
%multi-instanton contributions.
%The latter, however, have been derived by a very formal 
%analytic continuation with respect to the coupling constant.
%We think the meaning or the structure of the procedure is not
%clear  and  it needs a clarification.

Our analysis sheds a new light on this topic: First, it explains naturally 
why the Borel singularity of the perturbation theory is canceled 
by the non-perturbative contribution induced by instantons, or more
generally, valley-instantons.
The interplay of the perturbative  and
non-perturbative contribution can be understood by the decomposition of the
integral contour for the collective
coordinate of the valley.
Second, it predicts the large order behaviors of the perturbative 
series for the asymmetric double-well model, which interestingly defeats
widely held beliefs that the perturbative series would be
Borel-summable for states without quantum tunneling \cite{folk}.
The valley-instanton \cite{AKHSW, AKHOSW}
provides us with the basis for the path-integral analysis.
Some assumptions that underlie our analysis are tested 
by comparing our theoretical predictions with the numerical
and exact calculations of the perturbative series.

The model we study is a one-dimensional system with coordinate $q$,
with the following Hamiltonian;
\begin{eqnarray}
H&=&- \frac{1}{2}\, \frac{d^2}{dq^2}+V(q),\nonumber\\
V(q)&=&\frac{1}{2}\, q^2(1-g q)^2-\epsilon g q,  
\end{eqnarray}
where $g (>0)$ is a coupling constant and $\epsilon (>0)$ is a parameter which
represents the asymmetry of the potential.
For $\epsilon g^2 < \sqrt{3}/18 $, this potential has two minima. 
Although our consideration is not restricted to the special kind of
model, the reason we adopt the model is that it becomes supersymmetric at the
special values of $\epsilon$.
For $\epsilon=1, 2$, this model is known to be supersymmetric
\cite{Salo,BY, VWW}.
The non-renormalization theorem makes clear the relation between the
perturbative contribution and the non-perturbative one and gives a rich structure. 
We will see that the result of the valley analysis agrees with 
the predictions of the supersymmetry.

In the case of $\epsilon=0$, the simplest example of the valley represents
an instanton--anti-instanton pair.
It approaches to the
infinitely separated instanton--anti-instanton configuration at one
end and has a vacuum configuration at the other end;
namely, as the instanton and the anti-instanton approach each other, 
the action decreases smoothly,
and finally when they coincide, they annihilate each other 
and the configuration becomes the vacuum.
A precise definition of the valley is given by the valley 
equation \cite{BY2,AK}.
The valley equation which we adopt here is 
\begin{eqnarray}
\int d\tau'\frac{\delta^2 S}{\delta q(\tau)\delta q(\tau')}
\cdot\frac{\delta S}{\delta q(\tau')}=
\lambda\frac{\delta S}{\delta q(\tau)},  
\label{veqn}
\end{eqnarray}
where $S$ is the action and $\lambda$ is a constant \cite{AK}.
The configuration which satisfies this equation extremizes the norm of 
the gradient vector 
$\int d\tau (\delta S/\delta q)^2$ on the contour plane with $S$ fixed.
Then, if we identify the value of the action as the ``height'' of 
the configuration, this configuration is in a ``valley'' in the 
configuration space. 
The instanton--anti-instanton pair is not a classical solution, but it can be
defined as a solution of the valley equation \cite{AK}.
When instanton and anti-instanton are well-separated from each other, 
the collective
coordinates of the valley are the relative distance between them and the
center of mass. 

The power of the valley equation is demonstrated in the case of 
$\epsilon\neq0$.
For this case, there is no instanton-like classical solution
which starts from one minimum of the potential and ends at the other minimum.  
But such a solution exists for the valley equation (\ref{veqn}) with
$\lambda$ set to zero \cite{AKHSW}.
We dub the solution which starts from the (right) left minimum 
as the (anti-)valley-instanton.
When $\epsilon g^2$ is small, the valley-instanton for the present model 
may be constructed in a similar way as given in \cite{AKHSW};
\begin{eqnarray}
q(\tau)=\left\{
\begin{array}{ll}
\displaystyle
\epsilon g+\frac{1}{g}e^{\omega_{+}\tau}
+\frac{3\epsilon g}{\omega_{+}}\tau e^{\omega_{+}\tau}+\cdots
& \mbox{if $\tau\ll-1$ ;}\\  
\noalign{\vspace{2mm}} \displaystyle
\frac{1}{g}\frac{1}{1+e^{-\tau}}+\epsilon g+\cdots 
& \hspace{-5mm} 
\mbox{if $-1/\epsilon g^2 \ll \tau\ll 1/\epsilon g^2 $ ;}\\
\noalign{\vspace{2mm}} \displaystyle
\frac{1}{g}+\epsilon g
-\frac{1}{g}e^{-\omega_{-}\tau}-\frac{3\epsilon g}{\omega_{-}}\tau 
e^{-\omega_{-}\tau}+\cdots
& \mbox{if $\tau\gg 1$ ,}
\end{array}
\right.
\end{eqnarray}
where
$\omega_{\pm}=1\mp 3\epsilon g^2 + \cdots$.
When $\epsilon$ goes to zero, the (anti-)valley-instanton smoothly 
converges to the ordinary (anti-)instanton.
This gives a natural extension of the instanton--anti-instanton pair in
the case of $\epsilon\neq 0$.
%Since parity is explicitly broken in this case, 
There are two types of the valleys, 
one made of a valley-instanton--anti-valley-instanton pair and the 
other of an anti-valley-instanton--valley-instanton pair \cite{AKHSW}.
We call the former as I-A valley and the latter A-I valley.\footnote{
An interesting property of the I-A valley (A-I valley) is that 
it contains the bounce solution if $\epsilon>0$ ($\epsilon<0$).
In most theories, the bounce solution is a signal for instability, 
then the existence of the bounce solution in our stable system
was somewhat mysterious.
In our valley, the integration of the negative mode of the bounce turns 
out to be the integration of the collective coordinate of the valley which
corresponds to the relative distance between valley-instanton and
anti-valley-instanton and 
cause no instability in the energy spectrum \cite{AKHSW}.
}

Now let us evaluate the transition amplitude from the left minimum
$|0_L\rangle$ to itself
\begin{eqnarray}
Z=\lim_{T\rightarrow\infty}\langle 0_L|e^{-HT}|0_L\rangle,
\end{eqnarray}
in the background of the I-A valley.
For simplicity, we focus mainly on the case of $\epsilon=0$.
Except for the collective coordinate of the valley, 
the path integral can be performed by the Gaussian integral.
Then we obtain 
\begin{eqnarray}
Z=\lim_{T\rightarrow\infty}{\rm const.}\times
\frac{T}{\sqrt{g^2}}\int_{0}^{T}dR \, e^{-S(R)/g^2},  
\label{valley:eqn}
\end{eqnarray}
where $R$ is the collective coordinate of the valley corresponding to
the relative distance between the valley-instanton and 
the anti-valley-instanton, and $S(R)/g^2$ 
is the action of the valley.\footnote{When the instanton and
anti-instanton are not far enough apart from each other, 
there is no natural definition of $R$.
We define $R$ so as to give Eq.(\ref{valley:eqn}) and (\ref{action:eqn}) .
}
A factor $T/\sqrt{g^2}$ results from the integration of the collective
coordinate for the translational symmetry.
When $\epsilon=0$, the action of the valley behaves as \cite{AK}
\begin{eqnarray}
S(R)=\left\{
\begin{array}{ll}
\displaystyle 
\frac{R^2}{2}& \mbox{for $R\rightarrow 0$ ;}\\
\noalign{\vspace{2mm}} \displaystyle
\frac{1}{3}-2e^{-R}& \mbox{for $R\rightarrow\infty$.}
\end{array}
\right.
\label{action:eqn}
\end{eqnarray}
The valley at $R=0$ is the vacuum and at $R\sim\infty$ the
well-separated instanton--anti-instanton pair;
the term $1/3$ in Eq.(\ref{action:eqn}) is their action and $-2e^{-R}$
is the interaction between them.
If we change the integration variable $R$ to $t=S(R)$, 
the amplitude (\ref{valley:eqn}) becomes
\begin{eqnarray}
Z=\lim_{x\rightarrow 1/3}{\rm const.}\times\frac{T}{\sqrt{g^2}}
\int_{C_V}dt F(t) \, e^{-t/g^2},  
\label{cv:eqn}
\end{eqnarray}
where $x=S(T)$, $C_V=[0,x]$, and $F(t)$ is the Jacobian.
From Eq.(\ref{action:eqn}), it is found that the Jacobian behaves as
\begin{eqnarray}
F(t)=\left\{
\begin{array}{ll}
\displaystyle \frac{1}{\sqrt{2t}}& \mbox{for $t\rightarrow 0$ ;}\\
\noalign{\vspace{2mm}} \displaystyle \frac{1}{1/3-t}& \mbox{for $t\rightarrow 
\displaystyle \frac{1}{3}$ .}
\end{array}
\right.
\end{eqnarray}
To obtain the full form of the Jacobian, the detailed analysis of the valley 
is needed.
Instead, we simply assume that the full form of the Jacobian is given by
\begin{eqnarray}
F(t)=\frac{f(t)}{\sqrt{2t}\, (1/3-t)},  
\end{eqnarray}
where $f(t)$ is an analytic function that satisfies
\begin{eqnarray}
f(0)={1 \over 3},\quad f(1/3)=\sqrt{2 \over 3}.  
\end{eqnarray}
No more details of $f(t)$ are needed in the following analysis.

\begin{figure}
\centerline{\epsfxsize=9cm\epsfbox{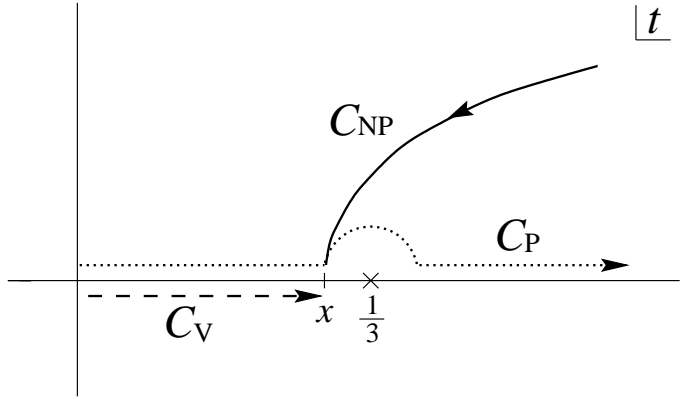}}
\caption{Deformation of  the contour $C_V$ 
to the sum of $C_{\rm P}$ and $C_{\rm NP}$.}
\label{fig:cv}
\end{figure}

The integral (\ref{cv:eqn}) contains both the perturbative contribution
at $t\sim 0$ and the non-perturbative one at $t\sim 1/3$.
To separate the perturbative and non-perturbative
contributions, we deform the contour $C_{\rm V}$ 
to the sum of $C_{\rm P}$ and $C_{\rm NP}$ as is shown in Fig.\ref{fig:cv}. 
Then the amplitude becomes 
\begin{eqnarray}
Z&=&\lim_{x\rightarrow 1/3}{\rm const.}\times\frac{T}{\sqrt{g^2}}
%Z&=& {\rm const.}\times\frac{T}{\sqrt{g^2}}
\int_{C_{\rm P}} dt F(t)\, e^{-t/g^2} \nonumber\\
&&  + \lim_{x\rightarrow 1/3}{\rm const.}\times
\frac{T}{\sqrt{g^2}}\int_{C_{\rm NP}} dt F(t)\, e^{-t/g^2}.  
\label{borel:eqn}
\end{eqnarray}
Note that there is a significant resemblance between the first term 
on the right-hand side and the formal Borel-summation of 
the perturbation series.
The singularity of $F(t)$ at $t=1/3$ is the same as that 
of the Borel function, which indicates the non-Borel-summability of the
perturbation series.
One of the present authors conjectured 
that this decomposition is essential to understand the 
interplay of the instanton effect to the perturbative calculation \cite{Kik}.
We identify that the first term of Eq.(\ref{borel:eqn}) is 
the formal Borel-summation of the perturbative series 
and second term is the non-perturbative one.
We show that this decomposition naturally reproduce 
all the known results of the interplay of them. 
We also present the predictions of the conjecture and test them.
We denote the first term of Eq.(\ref{borel:eqn}) as $Z_{\rm P}$ and the second
term as $Z_{\rm NP}$. 

An immediate consequence of the decomposition is 
\begin{eqnarray}
{\rm Im}Z_{\rm P}+{\rm Im}Z_{\rm NP}=0.  
\end{eqnarray}
This is because $Z=Z_{\rm P}+Z_{\rm NP}$ is real.
This simple equation explains why the imaginary part 
of the formal Borel-summation of the perturbation series is canceled 
by that of the instanton contribution. 
At the same time, it also shows that non-zero imaginary part of the
non-perturbative contribution is a
necessary and sufficient condition for the non-Borel-summability of the
perturbative expansion. 
Furthermore, when the imaginary part is not zero, 
we can predict the large order
behavior of the perturbative contribution from the dispersion relation.
To this end, we examine the analyticity of $Z_{\rm P}(g^2)$ in the
complex $g^2$-plane.
When the phase of $g^2$ changes to $2\pi$, the perturbative amplitude
$Z_{\rm P}(g^2)$ changes as
\begin{eqnarray}
Z_{\rm P}(g^2)\rightarrow Z_{\rm P}(g^2e^{2\pi i})
=Z_{\rm P}(g^2) - 2 i {\rm Im}Z_{\rm P}(g^2),
\end{eqnarray}
then, $Z_{\rm P}(g^2)$ has a cut on the real axis in the complex $g^2$
plane.
This is the only singularity near the origin. 
Thus the dispersion relation becomes
\begin{eqnarray}
Z_{\rm P}(g^2)&=&\frac{1}{\pi}\int_{0}^{\infty}dz
\frac{{\rm Im}Z_{\rm P}(z)}{z-g^2}+\cdots \nonumber\\
&=&-\frac{1}{\pi}\int_{0}^{\infty}dz\frac{{\rm Im}Z_{\rm NP}(z)}{z-g^2}
+\cdots,
\label{eqn:disp}
\end{eqnarray}
where we have neglected the contribution from the singularity far from
the origin.
If we formally expand the integrand, we obtain 
the following;\footnote{This result
holds for a sufficiently large $m$ even if $Z_{\rm P}(z)$ has power divergence
for $|z| \rightarrow \infty$ which necessitates subtractions 
for the dispersion relation Eq.(\ref{eqn:disp}).
We would like to thank Prof.~S.~Matsuda for bringing this point
to our attention.}
\begin{eqnarray}
Z_{\rm P}(g^2)&=&-\frac{1}{\pi}\sum_{m=0}^{\infty}
\int_{0}^{\infty}dz\frac{{\rm Im}Z_{\rm NP}(z)}{z^{m+1}}g^{2m}+\cdots.   
\label{zn:eqn}
\end{eqnarray}
%In Eqs.(\ref{eqn:disp}) and (\ref{zn:eqn}), we have neglected the other
%singularities that may appear in the integral; they are apart from the
%origin and does not give the leading contribution in the perturbative
%coefficients.
When $m$ is large enough, the coefficient of $g^{2m}$ is determined
by the singularity near the origin.
Then, we obtain the large order behavior of $Z_{\rm P}(g^2)$ as
\begin{eqnarray}
&&Z_{\rm P}(g^2)=\sum_{m=1}^{\infty}z_mg^{2m},\label{eqn:zpg}\\
&&z_m\stackrel{m\rightarrow\infty}{\sim}-\frac{1}{\pi}\int_0^{\infty}
dz\frac{{\rm Im}Z_{\rm NP}(z)}{z^{m+1}},\label{eqn:znlarge}
\end{eqnarray}
which reproduces the formula in \cite{Zin2}.

\begin{figure}
\centerline{\epsfxsize=9cm\epsfbox{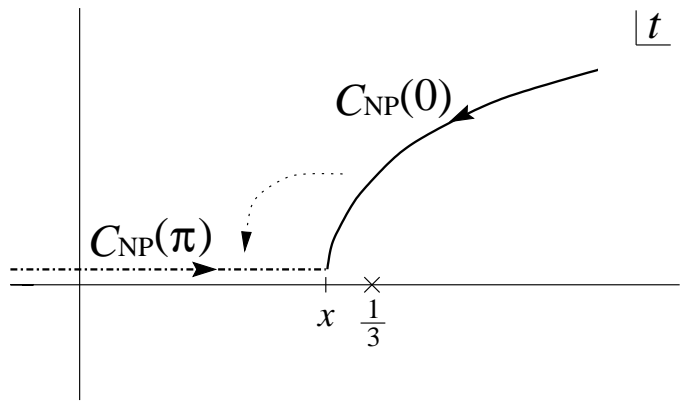}}
\caption{The change of the contour $C_{\rm NP}(\theta)$ 
%$C_{\rm NP}(0)$ to $C_{\rm NP}(\pi)$, 
as $\theta$ is changed from zero to $\pi$.}
\label{fig:cvrotation}
\end{figure}

We find that $Z_{\rm NP}$ also reproduces the old result for the
instanton--anti-instanton contribution as follows: 
For weak coupling, the dominate contribution of the integral of $Z_{\rm NP}$ 
comes from $t=1/3$. Then if we expand $F(t)$ as
\begin{eqnarray}
F(t)=-\frac{1}{t-1/3}+\cdots,   
\end{eqnarray}
and use $x=1/3-2e^{-T}+\cdots$, we obtain
\begin{eqnarray}
Z_{\rm NP}={\rm const.}\times\frac{T}{\sqrt{g^2}} \, e^{-1/3g^2}
\left[T-\ln(-2/g^2)-\gamma\right],  
\end{eqnarray}
where $\gamma$ is the Euler's constant.
This is the exactly same as the old result \cite{Bog}. 
An advantage of our viewpoint is that it clarifies the meaning of a
formal analytic continuation in the coupling constant \cite{Bog,Zin}.
For this purpose, we perform an analytic continuation for $Z_{\rm NP}(g^2)$.
We denote the contour for $Z_{\rm NP}(|g^2|e^{i\theta})$ as 
$C_{\rm NP}(\theta)$.
If $\theta$ is changed from zero to $\pi$, the contour
$C_{\rm NP}(0)$ changes to $C_{\rm NP}(\pi)$ (see Fig.\ref{fig:cvrotation}) and
in the weak coupling limit, $C_{\rm NP}(\pi)$ can be safely replaced
with $C_{\rm V}$.
The resultant integral coincides with that obtained by the formal analytic
continuation for $Z(g^2)$;  
$Z_{\rm NP}(|g^2|e^{i\pi})=Z(|g^2|e^{i\pi})$.
Therefore, in our viewpoint, the formal analytic continuation for
$Z(g^2)$ is naturally justified as the real analytic continuation 
for $Z_{\rm NP}(g^2)$.

For $\epsilon\neq 0$ case, although some complications arise, the main
result does not change. 
Especially, Eq.(\ref{eqn:zpg}) and (\ref{eqn:znlarge}) also hold in this case.

In general, there exist
valleys which connect the classical minima to the multi-valley-instantons.
These play important roles in the calculation of the spectrum of the
excited states. 
In the following, we incorporate them into the partition function,
$\lim_{T \rightarrow \infty} {\rm Tr} (e^{-HT})$,
which we also denote by $Z$.
We have not yet completed the analogous analysis of the I-A valley, 
but the present approach is sufficient to evaluate
the non-perturbative part $Z_{\rm NP}$ in the weak coupling limit.

When all the valley-instantons are well separated from each other,
the action of the multi-valley-instantons is derived by the standard
technique \cite{Bog,AK2}. 
If the valley consists of $n$ pairs of the (anti-)valley-instantons, we obtain
\begin{eqnarray}
S=\frac{n}{3g^2}-\epsilon \sum_{i=1}^{n}R_i-\frac{2}{g^2}
\sum_{i=1}^{n}e^{-R_i}-\frac{2}{g^2}\sum_{i=1}^{n}e^{-\tilde{R_i}},
\end{eqnarray}
where $R_i$ is the distance between the $i$-th 
valley-instanton and the $i$-th anti-valley-instanton 
and $\tilde{R_i}$ is that between
the $i$-th anti-valley-instanton and the $(i+1)$-th valley-instanton.
When $g^2=|g^2|e^{i\pi}$ and $|g|^2\ll 1$, the following equation is
meaningful since the interaction between the valley-instanton and the
anti-valley-instanton is repulsive in this case.
(Note that the interaction $\epsilon\sum R_i$ is not attractive since 
$\epsilon\sum R_i =\epsilon T-\epsilon\sum\tilde{R}_i$. 
See Eq.(\ref{eqn:22}) below.)
\begin{eqnarray}
Z=\lim_{T\rightarrow\infty}\sum_{n=1}^{\infty}\alpha^{2n}J_{n}(T),   
\label{eqn:Z}
\end{eqnarray}
where $\alpha=e^{-1/6g^2}/g\pi^{1/2}$, and 
\begin{eqnarray}
J_{n}(T)&=&\frac{T}{n}\int_{0}^{\infty}\prod_{i=1}^{n}dR_i 
%\prod_{i=1}^{n}
d\tilde{R}_i
\, \delta\left(\sum_{i=1}^{n}(R_{i}+\tilde{R}_i)-T\right)\nonumber\\ 
&& \times \exp\left(\epsilon\sum_{i=1}^{n}R_i+\frac{2}{g^2}
\sum_{i=1}^{n}e^{-R_i}+\frac{2}{g^2}\sum_{i=1}^{n}e^{-\tilde{R}_i}\right).
\label{eqn:22}
\end{eqnarray}
One lesson to draw from the analysis of the I-A valley is that 
$Z$ is equal to the non-perturbative amplitude $Z_{\rm NP}$
when $g^2=|g^2|e^{i\pi}$ and $|g^2|\ll 1$.
Turning back to $\theta=0$ after the evaluation of $Z$, 
we obtain $Z_{\rm NP}$ for positive $g^2$.

We extend the Zinn-Justin method \cite{Zin} to evaluate Eq.(\ref{eqn:Z}).
If we rewrite the delta function as
\begin{eqnarray}
\delta\left(\sum_{i=1}^{n}(R_i+\tilde{R}_i)-T\right)=\frac{1}{2\pi i}
\int_{-i\infty-\eta}^{i\infty-\eta}ds 
\exp\left(-sT+s\sum_{i=1}^{n}(R_i+\tilde{R}_i)\right),  
\end{eqnarray}
the integrals over $R_i$ and $\tilde{R}_i$ are factorized. After
summation over multi-valley-instanton contributions and turning back to
$\theta=0$, we finally obtain
\begin{eqnarray}
Z_{\rm NP}=-\lim_{T\rightarrow\infty}\frac{1}{2\pi i}
\int_{-i\infty-\eta}^{i\infty-\eta}ds \, e^{-Ts}\frac{\phi'(s)}{\phi(s)}, 
\end{eqnarray}
where
\begin{eqnarray}
\phi(s)&=&1-\alpha^2\int_{0}^{\infty}dR \, d\tilde{R} \, 
\exp\left((s+\epsilon)R+\frac{2}{g^2}e^{-R}
+s\tilde{R}+\frac{2}{g^2}e^{-\tilde{R}}\right)\nonumber\\
&=&1-\alpha^2\left(-\frac{2}{g^2}\right)^{2s+\epsilon}
\Gamma(-s-\epsilon)\Gamma(-s).
\label{master:eqn}
\end{eqnarray} 
If we denote the poles corresponding to the solutions of the equation
$\phi(s)=0$ as $s_n$, we obtain
\begin{eqnarray}
Z_{\rm NP}\sim\sum_{n} e^{-s_nT}.
\end{eqnarray}
Therefore, $s_n$ give the
non-perturbative contribution to the energy levels.

For small coupling, $s_n$ can be obtained as a perturbative series in $\alpha$.
For $\epsilon$ away from integer values, it yields the following
to the first nontrivial order, $\alpha^2$;
\begin{eqnarray}
E_{\rm NP}^{(+)}(\epsilon, N) &=& \frac{1}{2}+N+
a^{(+)}(\epsilon, N)\alpha^2,\nonumber\\
E_{\rm NP}^{(-)}(\epsilon, N) &=& -\epsilon + \frac{1}{2}+N+a^{(-)}
(\epsilon, N)\alpha^2,
\label{eqn:enp}
\end{eqnarray}
where
\begin{equation}
a^{(\pm)}(\epsilon, N)= \frac{(-1)^{N+1}}{N!}\Gamma(\mp\epsilon-N)
\left(-\frac{2}{g^2}\right)^{\pm\epsilon+2N}.
\end{equation}
The energies (\ref{eqn:enp}) are for the localized states in the left well 
and the right well, respectively, as can be seen in the 
free limit, $\alpha \rightarrow 0$.
For $\epsilon=N_0 (=0,1,2,\cdots)$, 
only $E_{\rm NP}^{(-)}(\epsilon, N)$ for $N \le N_0$ 
in the above is valid.  The rest has to be solved taking into account
the confluence of the poles of the two $\Gamma$-functions in 
Eq.(\ref{master:eqn}), which yields the following to the order $\alpha^2$;
\begin{eqnarray}
&&E_{\rm NP}(N_0, N, \pm)=\frac{1}{2}+N\pm\alpha\sqrt{\frac{1}{N!\, (N+N_0)!}
\left(\frac{2}{g^2}\right)^{2N+N_0}}\nonumber\\
&& \quad {}+{\alpha^2 \over 2}\left(\frac{2}{g^2}\right)^{2N+N_0}
\hspace{-9pt} \frac{1}{N!\, (N+N_0)!}
\left(2\ln\left(-\frac{2}{g^2}\right)+2\gamma
-\sum_{n=1}^{N}\frac{1}{n}-\sum_{n=1}^{N+N_0}\frac{1}{n}\right).\nonumber\\  
\label{eqn:dege}
\end{eqnarray}
The plus and the minus signs in the above expression Eq.(\ref{eqn:dege})
correspond to two linear combinations of the perturbative states in the
left and right well with the same zero-th order energy.
This situation is analogous to the lifting of the degeneracy by
the instanton contribution for the symmetric double-well potential.

The expression of the nonperturbative contribution to the energy
levels, Eq.(\ref{eqn:enp}) -- Eq.(\ref{eqn:dege}), contain
imaginary parts, which are of order $\alpha^2$ and are continuous
at integer $\epsilon$'s.  Using Eq.(\ref{eqn:znlarge}),
these imaginary parts lead to the following leading term
of the $m$-th order perturbative coefficient of the $N$-th level, 
$E_{\rm P}^{(\pm)}(\epsilon,N,m)$ 
($E_{\rm P}^{(\pm)}(\epsilon,N) = 
\sum_{m=0}^\infty E_{\rm P}^{(\pm)}(\epsilon,N,m) g^{2m}$);
\begin{eqnarray}
E_{\rm P}^{(\pm)}(\epsilon,N,m) &=& A^{(\pm)}
(\epsilon, N) 3^m \Gamma(\pm \, \epsilon + 2N + m +1) 
\left[1 + O\left( {1 \over m}\right)\right],\label{eqn:eaep}\\
A^{(\pm)}(\epsilon, N) &\equiv&
-{3 \over \pi} {6^{\pm\epsilon + 2N} \over 
N! \, \Gamma(\pm \, \epsilon + 1 + N)}.
\label{eqn:aep}
\end{eqnarray}
The expression for $E_{\rm P}^{(-)}(\epsilon, 1)$
coincides with the expression obtained in\break
Ref.\cite{VWW,VWW2}.

\begin{figure}
\centerline{\epsfxsize=9cm\epsfbox{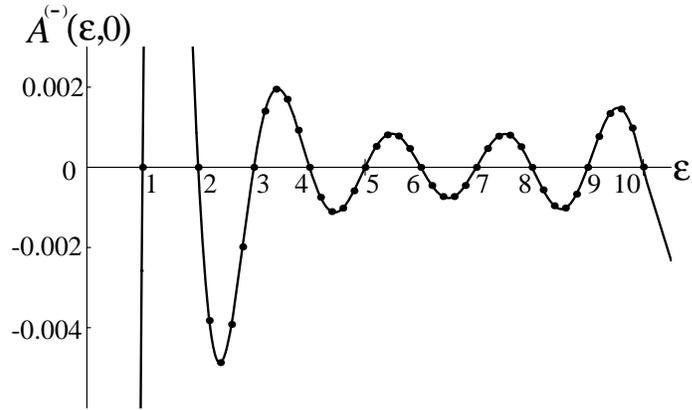}}
\caption{Comparison of the fitting to the perturbative coefficients 
(indicated by dots) and
the theoretical prediction (solid line) of $A^{(-)}(\epsilon, 0)$.}
\label{fig:aeps}
\end{figure}

We have independently carried out the numerical and exact 
calculation of the perturbative coefficients 
$E_{\rm P}^{(\pm)}(\epsilon, N, m)$
by the methods described in Ref.\cite{VWW,Zin2,BPZtwo}
to the 200-th order for the following four categories;
(a) $N=0$ ($-$) level (ground state) for  $\epsilon = 0$
to 10 with $\Delta\epsilon=0.2$ interval, 
(b) $N=0$ ($+$) level for  $\epsilon = 0$ to 20 
with $\Delta\epsilon=0.2$,
(c) $N=3$ ($+$)-level up to $\epsilon=6.5$ with $\Delta\epsilon=0.5$,
(d) $N=1$ to 6 ($-$) levels for $\epsilon=2.5$.
In order to check the leading $m$-dependent terms in Eq.(\ref{eqn:eaep}),
we take their ratio,
\begin{equation}
{E_{\rm P}^{(\pm)}(\epsilon, N, m) \over E_{\rm P}^{(\pm)}(\epsilon, N,m-1)}
=3(\pm \, \epsilon + m).
\end{equation}
The agreement between the numerical fitting of 
the perturbative coefficients for $m=150$ to 200
and this theoretical prediction is excellent in all the calculated 
cases, with the maximum error of order 0.1\%.
Next we have calculated $A^{(\pm)}(\epsilon, N)$ defined by 
Eq.(\ref{eqn:eaep}) numerically and compared with Eq.(\ref{eqn:aep}).
The result for the case (a) is plotted in Fig.\ref{fig:aeps}.
The difference between Eq.(\ref{eqn:aep}) and the calculated value 
is at most 0.1 \% (at $\epsilon=9.8$).  
For the case (b), the error is at most 15 \% (at $\epsilon=20$).
For the cases (c) and (d), the maximum error is 0.15 \%.
In summary, the agreement between the expression (\ref{eqn:eaep}),
(\ref{eqn:aep}), 
and the actual perturbative coefficients is excellent in all the
cases examined.

\begin{figure}
\centerline{\epsfxsize=9cm\epsfbox{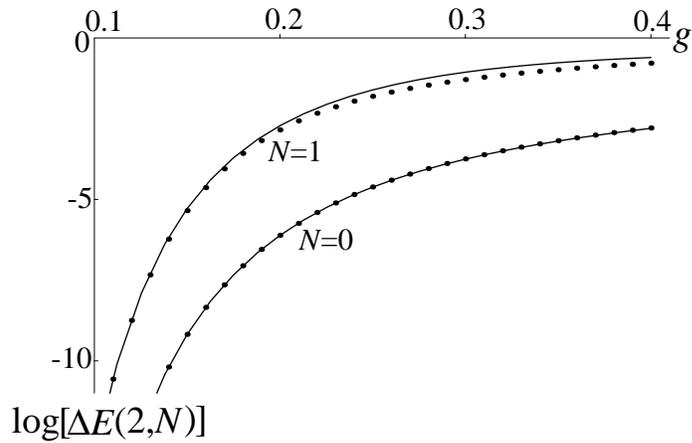}}
\caption{Comparison of our valley results and the numerical results
for the energy levels of the ground state and the first excited
states at $\epsilon=2$.  $\Delta E$ is the difference between
the energy level and the zero-th order results.
The solid line is our result, while the dots represent the 
numerical results.}
\label{fig:eptwo}
\end{figure}

The reader may note that  $A^{(-)}(\epsilon, 0)$ is zero for 
any positive integer $\epsilon$.  For $\epsilon=1$, this is because
of the supersymmetry, which prohibits any perturbative correction
to the ground state energy \cite{Salo}. 
For $\epsilon=2$, there are no perturbative corrections 
to the energy levels of the ground state and the
first excited state due to the supersymmetry \cite{VWW}.
We have calculated these energy levels numerically and compared them
with our valley result,  Eq.(\ref{eqn:enp}) -- Eq.(\ref{eqn:dege}).
The result is plotted in Fig.\ref{fig:eptwo} and the agreement 
is excellent.
For $\epsilon=3,4,\cdots$, there are
analogues  of supersymmetry that explain vanishing of 
$A^{(-)}$ and also some other results for excited states.  
These results will be published in near future. 

We stress that the bounce solution plays no important roles 
in the large order behavior of the perturbative series.
As a result of this, 
states that have no associated quantum tunneling phenomena do not
necessarily have Borel-summable perturbative series, in 
contrast to some beliefs \cite{folk}.
In fact, non Borel-summability is found for low-lying stable states 
when $\epsilon \ne$ integer, which is confirmed by the calculation
of the perturbative coefficients.

We believe that our analysis clearly shows that the valley 
is essential for the definition of the so-called ``non-perturbative
effects''; only when it is defined in the light of the separation from
the valley, it becomes physically sensible and calculable.
Same can be said of the perturbation theory;
the origin of its Borel singularity and its cancellation 
by the non-perturbative effects become evident in view of the valley.

\vspace{12pt}

H. Aoyama's work is supported in part by the Grant-in-Aid 
for Scientific Research (C)-07640391 and 09226219.
H.~Kikuchi's work is supported in part by the Grant-in-Aid 
for Scientific Research 09226232.
M.~Sato and S.~Wada's work is supported in part by the Grant-in-Aid 
for JSPS fellows.
Numerical computation in this work was in part supported by
the Yukawa Institute for Theoretical Physics.

%%%%%  Definitions %%%%%%%%%%%%%%%%%%%%%%%%%%%%%%%%%%%%%%%%%%%%%%%%%%%%%
\newcommand{\J}[4]{{\sl #1} {\bf #2} (19#3) #4}
\newcommand{\MPL}{Mod.~Phys.~Lett.}
\newcommand{\NP}{Nucl.~Phys.}
\newcommand{\PL}{Phys.~Lett.}
\newcommand{\PR}{Phys.~Rev.}
\newcommand{\PRL}{Phys.~Rev.~Lett.}
\newcommand{\AP}{Ann.~Phys.}
\newcommand{\CMP}{Commun.~Math.~Phys.}
\newcommand{\CQG}{Class.~Quant.~Grav.}
\newcommand{\PRP}{Phys.~Rept.}
\newcommand{\SPU}{Sov.~Phys.~Usp.}
\newcommand{\RMPA}{Rev.~Math.~Pur.~et~Appl.}
\newcommand{\SPJ}{Sov.~Phys.~JETP}
\newcommand{\MP}{Int.~Mod.~Phys.}
\newcommand{\JMP}{J.~Math.~Phys.}
%%%%  Contents %%%%%%%%%%%%%%%%%%%%%%%%%%%%%%%%%%%%%%%%%%%%%%%%%%%%%%%%%

\newpage

\end{document}